\documentclass[12pt,preprint]{aastex}

\shorttitle{Solar coronal magnetic field}

\shortauthors{Sasikumar Raja \& Ramesh}

\begin{document}

\title{Low frequency observations of transient 
quasi-periodic radio emission from the solar atmosphere}

\author{K. Sasikumar Raja\altaffilmark{1} and R. Ramesh\altaffilmark{1}}
\email{sasikumar@iiap.res.in}

\altaffiltext{1}{Indian Institute of Astrophysics, II Block, Koramangala, Bangalore - 560 034.}

\begin{abstract}

We report low frequency observations of 
the quasi-periodic, circularly polarized, harmonic type III radio bursts 
whose associated sunspot active regions
were located close to the solar limb. The measured periodicity of the bursts 
at 80 MHz was $\approx$ 5.2 s and
their average degree of circular polarization ($dcp$) was $\approx 0.12$. 
We calculated the associated magnetic field $B$ : (1) using the empirical relationship between 
the $dcp$ and $B$ for the harmonic type III emission, and (2) from the observed quasi-periodicity of 
the bursts. Both the methods result in $B \approx$ 4.2 G at
the location of the 80 MHz plasma level (radial distance $r \approx 1.3~\rm R_{\odot}$) in the active region corona. 

\end{abstract}


\section{Introduction}

Type III solar radio bursts are the signatures of accelerated electrons
streaming outward through the corona and the interplanetary medium in the aftermath of
flares and weak chromospheric brightenings. 
The electrons stream at speeds $\approx c/3$ along the open field magnetic field
lines. The radio emission is widely accepted to be 
because of the following two-step process: (1) the
excitation of high levels of plasma oscillations (Langmuir waves) by the propagating 
electron streams/beams and (2) subsequent conversion of these Langmuir waves into 
electromagnetic waves
at the fundamental (F) and second harmonic (H) of the local plasma frequency. 
The H emission results from the coalescence of two Langmuir waves. 
The type III
bursts are best identified on the 
dynamic spectra obtained using radio spectrographs.
They appear as intense emission, drifting from 
high to low frequencies at $\approx$ 100 MHz/s. 
The above frequency drift results from the decrease of
electron density and hence the plasma frequency, with distance 
in the solar atmosphere.
The bursts occur over the frequency range from $\approx$ 1 GHz-10 kHz, corresponding
to a distance range extending from the low corona to beyond the orbit
of the Earth. 
The bursts occur either isolated or in groups.
While the former has a life-time of about few seconds at any given frequency, 
the groups of bursts last for $\approx$ 5 m \citep{Suzuki1985}.
Groups of type III bursts are generally associated with flares 
observed in X-rays and/or H$\alpha$. They are a classic
signature of the impulsive phase of flares in which the radiation 
is primarily non-thermal. It is believed that each individual 
burst is due to radiation associated with electron streams moving outwards
through the corona along large scale diverging magnetic field lines 
connected to a common acceleration/injection site \citep{Mercier1975,Pick1986}.
The occurence
of non-thermal bursts is a clear indication of particle acceleration 
during an event. Note that the generation of radio emission by
non-thermal electron beams is very efficient due to the coherent 
nature of the emission mechanism.
Most of the type III bursts are weakly, circularly polarized with $dcp <$ 0.15. 
However, some
bursts, those identified as due to fundamental plasma emission, have 
$dcp\approx$ 0.5 \citep{Suzuki1978,Dulk1980}. Harmonic emission is observed mainly
at low frequencies. 
Note that the presence of the background magnetic field at the
source region of type III radio bursts can give rise to a net $dcp$ in
the ordinary (`o') mode for the escaping radiation \citep{Melrose1972,
Melrose1978,Melrose1980,Zlotnik1981,Willes1997}. 
The polarization observations of the H component of
type III radio bursts are considered to be a better diagnostic tool
to estimate the solar coronal magnetic field since the polarization
of the F component is affected by propagation effects \citep{Dulk1978}. 
In this paper we report observations of quasi-periodic, harmonic type III bursts 
\citep{Wild1963,Janssens1973,Mangeney1989,Zhao1991,Aschwanden1994,Ramesh2003,Ramesh2005}
and estimate the coronal magnetic field using the empirical relationship 
between the magnetic field and the $dcp$ for 
the harmonic plasma emission \citep{Melrose1978}. We verified the results by independantly 
estimating the
magnetic field from the periodicity of the observed type III radio bursts. 
Interestingly both the methods give consistent results indicating the usefulness of 
ground based low frequency radio observations to estimate the coronal magnetic field, 
particularly over the range $1.2~R_{\odot} \lesssim r \lesssim 2~R_{\odot}$, where 
measurements at other wavelength bands in the electromagnetic spectrum are 
presently difficult.

\section{Observations}

The radio data reported in the present work were obtained at 80 MHz on
2012 September 20, 2013 January 18, and 2013 March 11 with the 
heliograph \citep{Ramesh1998,Ramesh1999b,Ramesh2006}, 
the polarimeter \citep{Ramesh2008}, and the
spectrograph \citep{Ebenezer2007,Kishore2013} at the
Gauribidanur observatory\footnote{http://www.iiap.res.in/centers/radio},
about 100 km north of Bangalore in India \citep{Ramesh2011a}.
The co-ordinates
of the array are Longitude = $77^{\arcdeg}27^{\arcmin}07^{\arcsec}$ east,
and Latitude = $13^{\arcdeg}36^{\arcmin}12^{\arcsec}$ North.
The heliograph (Gauribidanur RAdioheliograPH, GRAPH) is a T-shaped
radio interferometer array which produces two 
dimensional images of the solar corona with an 
angular resolution of 
$\approx 5^{\arcmin} \times 7^{\arcmin}$ ($\rm R.A. \times decl.$)
at the above frequency. The integration time is $\approx$ 250 ms
and the observing bandwidth is $\approx$ 2 MHz.
The polarimeter (Gauribidanur Radio Interference Polarimeter, GRIP) 
is an east-west one-dimensional array operating in the
interferometer mode and it
responds to the integrated and polarized flux densities  
from the `whole' Sun. 
The half-power width of the response function (`beam') of the GRIP
is broad (compared to the Sun) in both 
right ascension/east-west direction ($\approx 2^{\arcdeg}$ 
at 80 MHz)
and declination/north-south direction
($\approx 90^{\arcdeg}$). This implies that a plot of the
GRIP data (i.e. the time profile) for observations
in the transit mode 
is essentially the `east-west beam' 
of the array with an amplitude proportional to the strength of the
emission from the `whole' Sun 
at the observing frequency, weighted by the antenna gain
in that direction.
The integration time is $\approx$ 250 ms
and the observing bandwidth is $\approx$ 2 MHz.
The radio source(s) responsible for the 
circularly polarized emission observed with the GRIP 
are identified using the two-dimensional
radioheliograms obtained with the GRAPH around the same
time.
Note that linear polarization, if
present at the coronal source region, tends to be obliterated
at low radio frequencies because of the differential Faraday
rotation of the plane of polarization within the observing
bandwidth \citep{Grognard1973}.
The spectrograph antenna system 
(Gauribidanur LOw frequency Solar Spectrograph, GLOSS) 
is a total power instrument and the half-power width of its
antenna response of GLOSS is 
$\approx 90^{\arcdeg} \times 6^{\arcdeg}$ ($\rm R.A. \times decl.$)
at 80 MHz. 
The integration time is $\approx$ 100 ms
and the observing bandwidth is $\approx$ 300 KHz at
each frequency.
The width of the response of GLOSS in hour
angle is nearly independent of frequency. The Sun is a
point source for both the GRIP and the GLOSS.
All the above three instruments observe the Sun everyday
during the interval 4 - 9 UT. 
The minimum detectable flux
density of the GRAPH, GRIP and GLOSS are
$\approx$ 20 Jy, 200 Jy, 3000 Jy, respectively
(1 Jy = $\rm 10^{-26}~Wm^{-2}Hz^{-1}$).

\begin{figure}
\epsscale{0.80}
\plotone{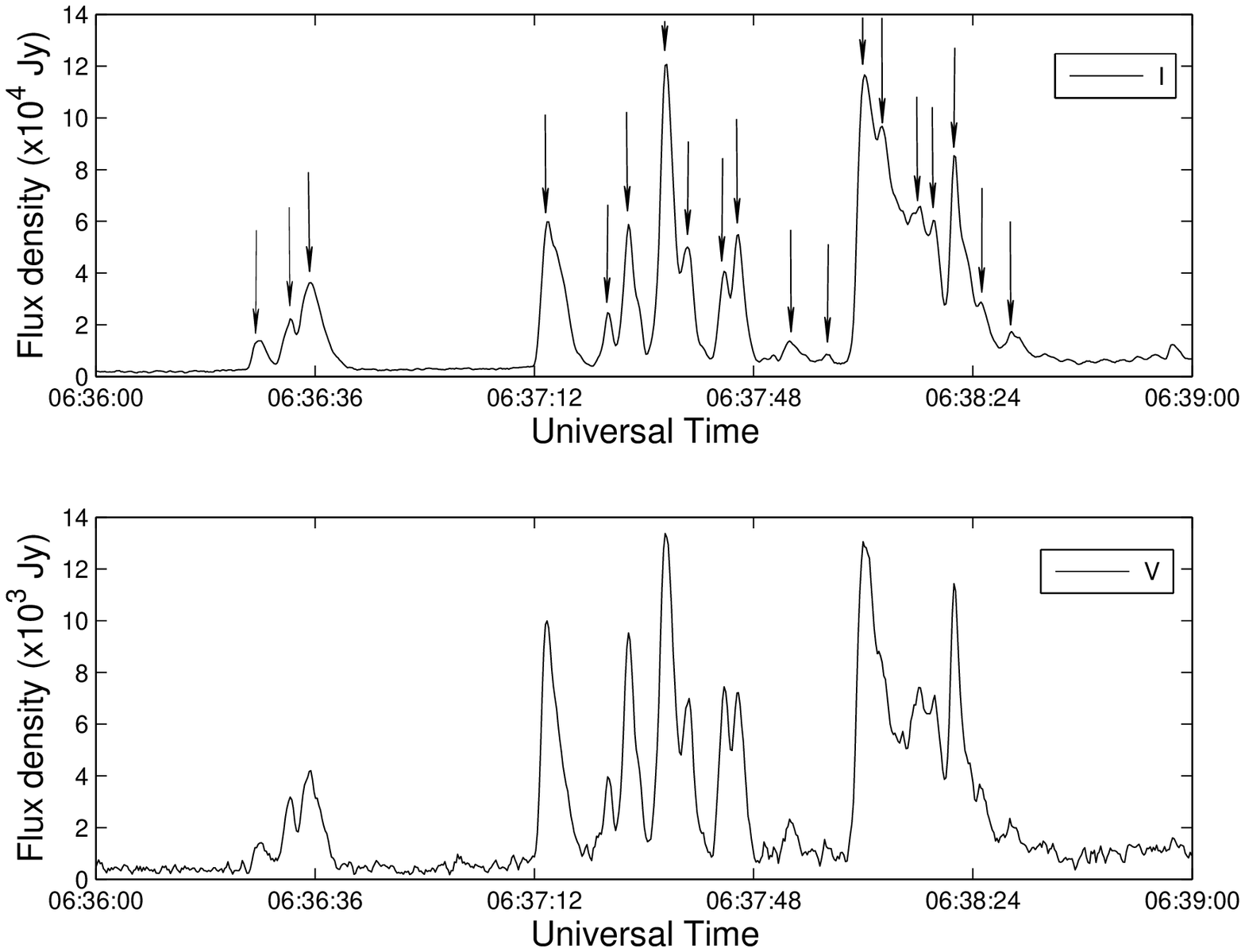}
\caption{GRIP observations of group of type III solar radio bursts on 
2012 September 20 at 80 MHz  
in Stokes I (upper panel) and Stokes V (lower panel). 
The arrow marks indicate the individual bursts.}
\label{fig:figure1}
\end{figure}

\begin{figure}
\epsscale{0.80}
\plotone{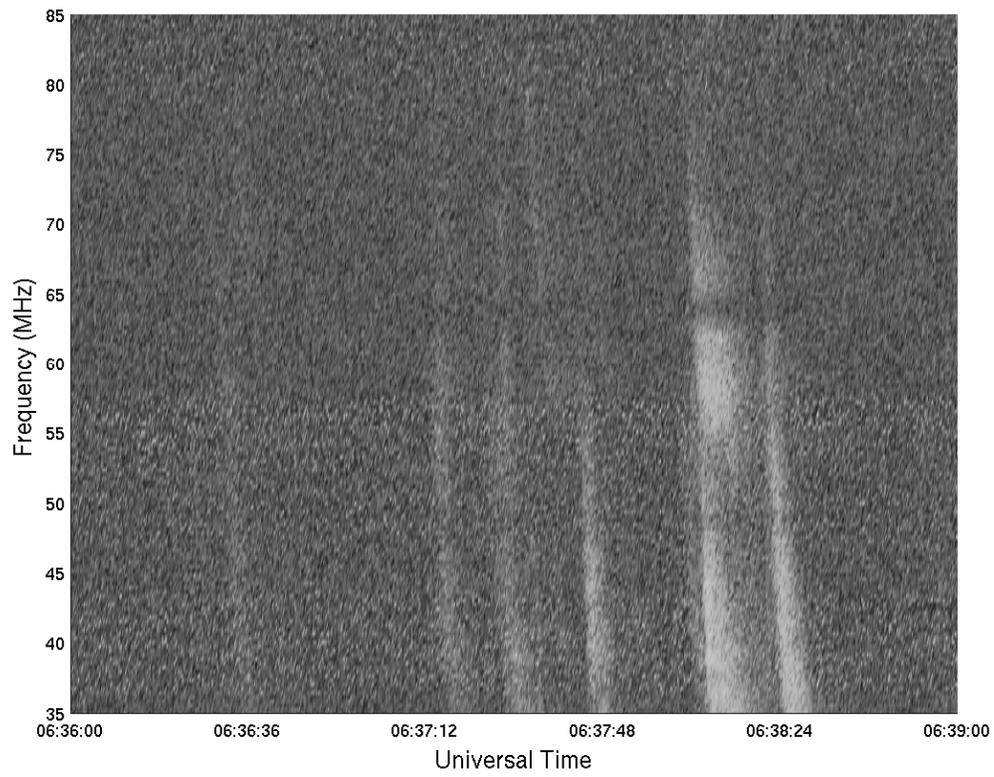}
\caption{GLOSS dynamic spectra (85-35 MHz) of the group of type III solar radio bursts in
Figure \ref{fig:figure1}.}
\label{fig:figure2}
\end{figure}

\begin{figure}
\epsscale{0.80}
\plotone{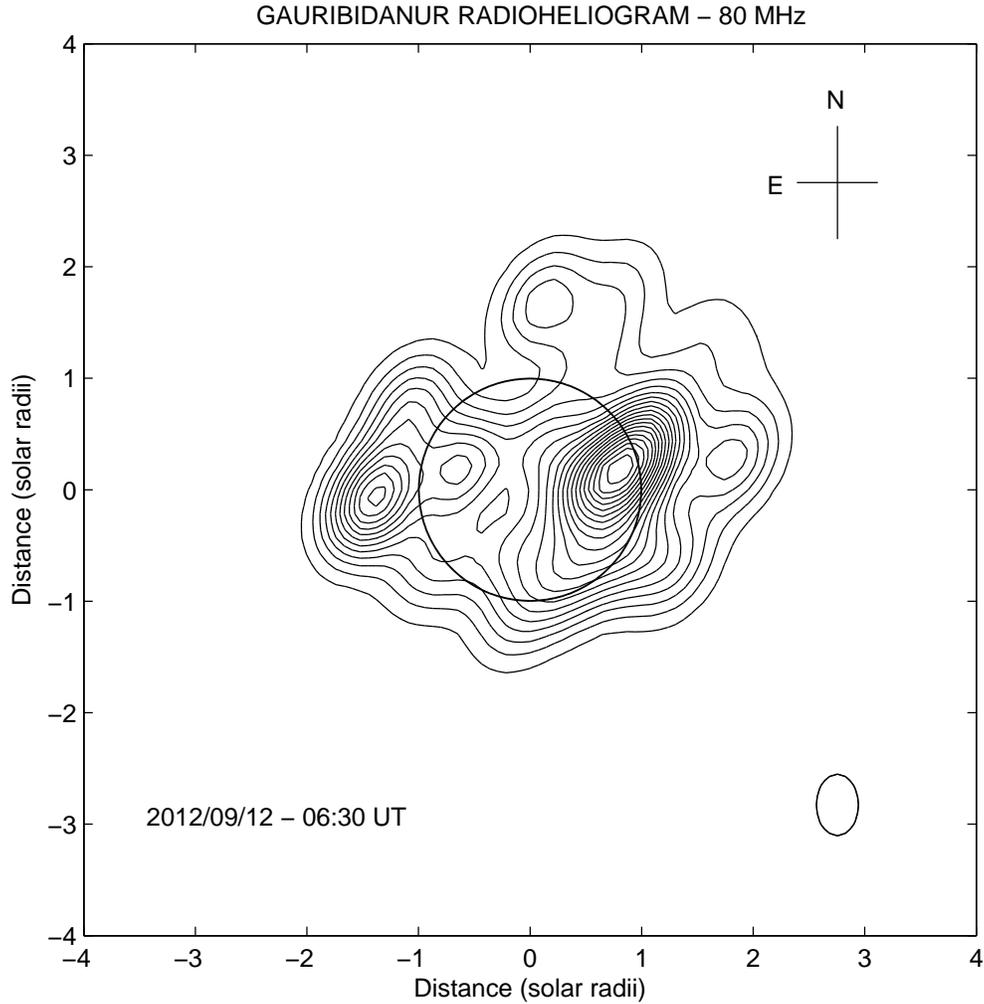}
\caption{GRAPH radioheliogram obtained on 2012 September 20 around 06:30 UT, prior to the
quasi-periodic type III burst emission in Figure \ref{fig:figure1}. The open circle at the
center represents the solar limb. The size of the GRAPH beam at 80 MHz is shown near
the lower right. The intense discrete source close to the west limb is the source region 
of the type III bursts in Figures \ref{fig:figure1} and \ref{fig:figure2}.} 
\label{fig:figure3}
\end{figure}

\begin{figure}
\epsscale{0.80}
\plotone{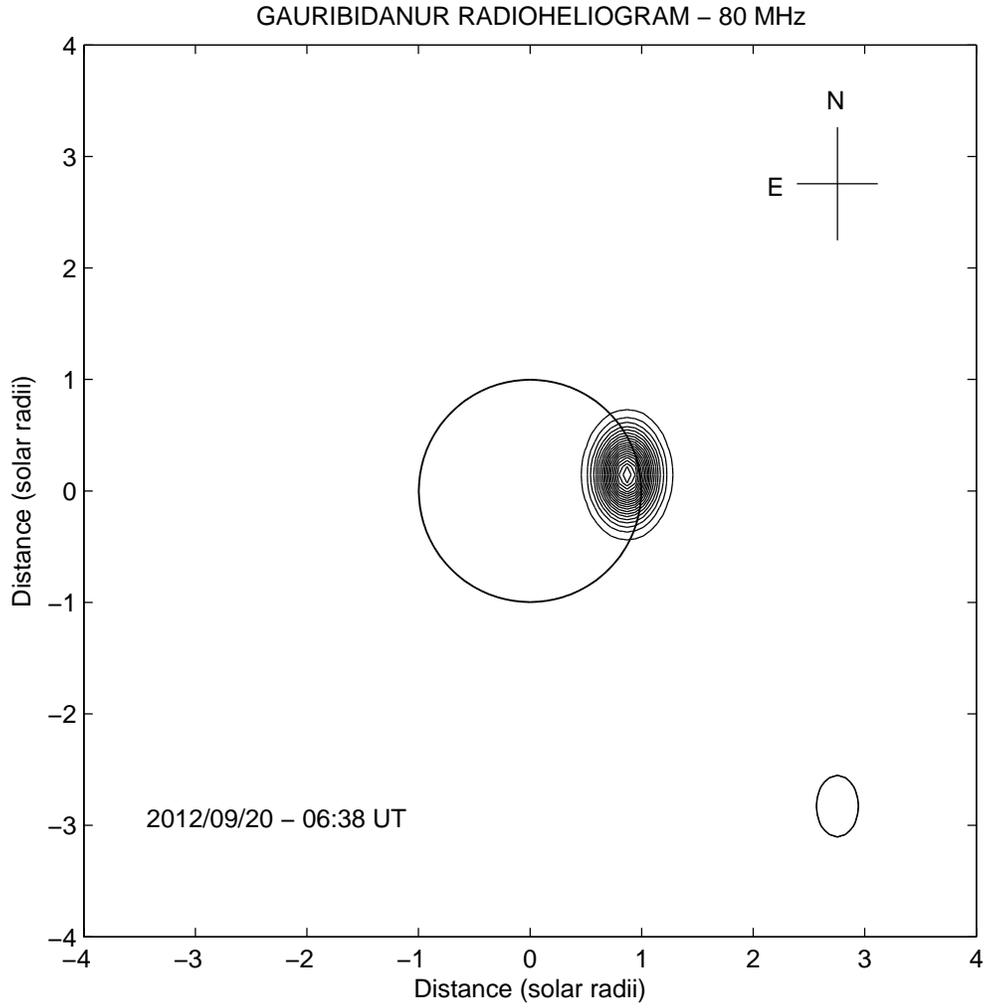}
\caption{Same as Figure \ref{fig:figure3} but obtained during the peak phase
of the burst in Figure \ref{fig:figure1} at $\approx$ 06:38 UT.}
\label{fig:figure4}
\end{figure}

Figure \ref{fig:figure1} shows
the temporal evolution of the Stokes I \& V radio emission 
from the solar corona at 80 MHz as observed with the GRIP 
on 2012 December 20 during the interval 06:36-06:39 UT, i.e. around the
transit of the Sun over Gauribidanur. One can notice 
the presence of intense quasi-periodic emission in both the 
Stokes I and V time profiles. These are the
characteristic signature of groups of type III solar 
radio bursts from the Sun \citep{Suzuki1985}.
We verified the same from the dynamic spectra (85-35 MHz) obtained with 
GLOSS during the above interval (Figure \ref{fig:figure2}).
The peak Stokes I and V flux densities in Figure \ref{fig:figure1} 
are $\approx 1.2 \times 10^{5}$ Jy and 
$1.6 \times 10^{4}$ Jy, respectively. The corresponding 
$dcp$ = $\rm \frac{|V~flux|}{I~flux}$ is 
$\approx$ 0.11. 
Note the average flux density of the type III solar radio
bursts reported in the literature is 
typically $\approx 10^{7}$ Jy \citep{Suzuki1985}. 
This is about two orders of magnitude higher than the 
peak flux density of the type III burst of 2012 December 20,
indicating that the latter is a weak event.
The estimated average periodicity of the 
radio emission in Figure \ref{fig:figure1} is $\approx$ 4.5 sec,
in both Stokes I \& V. The source region of the bursts 
was identified from the radioheliogram obtained with
the GRAPH around the same time (Figures \ref{fig:figure3}
and \ref{fig:figure4}).
One can notice the bursts orginated close to the solar limb.
They were associated with a SF class H$\alpha$ flare from 
AR 11574\footnote{http://www.swpc.noaa.gov/} 
located at S25W69\footnote{http://www.lmsal.com/solarsoft/latest\_events/}.
A shift in the solar radio source position due to ionospheric
effects is expected to be $\lesssim 0.1~R_{\odot}$ at 80 MHz in 
the hour angle range $\pm$ 2 hr \citep{Stewart1982}. The 
present observations were carried out close to the 
transit of the Sun over the local meridian at Gauribidanur.
Similarly, the effects of
scattering (irregular refraction due to density inhomogeneities
in the solar corona) on the observed source position are
also considered to be small at 80 MHz
compared to lower frequencies \citep{Aubier1971,Bastian2004}. 
The positional shift of discrete solar radio sources due to scattering
is expected to be $\lesssim 0.2~R_{\odot}$ at 80 MHz 
\citep{Riddle1974,Robinson1983,Thejappa2007}.
Ray tracing calculations employing 
realistic coronal electron
density models and density fluctuations show that the turning
points of the rays that undergo irregular refraction almost
coincide with the location of the plasma (`critical') layer
in the non-scattering case even at 73.8 MHz \citep{Thejappa2008}. 
Obviously, the situation should be better at 80 MHz. 
Also there is no unambiguous evidence that scattering is important
in any of the solar radio bursts \citep{McLean1985}.
Note that high angular resolution 
observations of the solar corona  
indicate that discrete radio sources of angular size $\approx 1^{\arcmin}$ 
are present in the solar atmosphere from where low frequency 
radio radiation originates \citep{Kerdraon1979,Lang1987,
Willson1998,Ramesh1999b,Ramesh2000,Ramesh2001,Mercier2006,Kathiravan2011,
Ramesh2012}. The projection effects are also expected to be
minimal since we have used only the limb events for the
present work (see for eg. Figure \ref{fig:figure4}).
The details related to the type III radio burst 2012 December 20 
described above and the other two events (2013 January 18 and 2013 March 11) 
are listed in Table \ref{tab:table1}. 

\section{Analysis and Results}

\subsection {Estimate of $B$ from the estimated $dcp$ of the type III radio bursts}

\begin{figure}
\epsscale{0.80}
\plotone{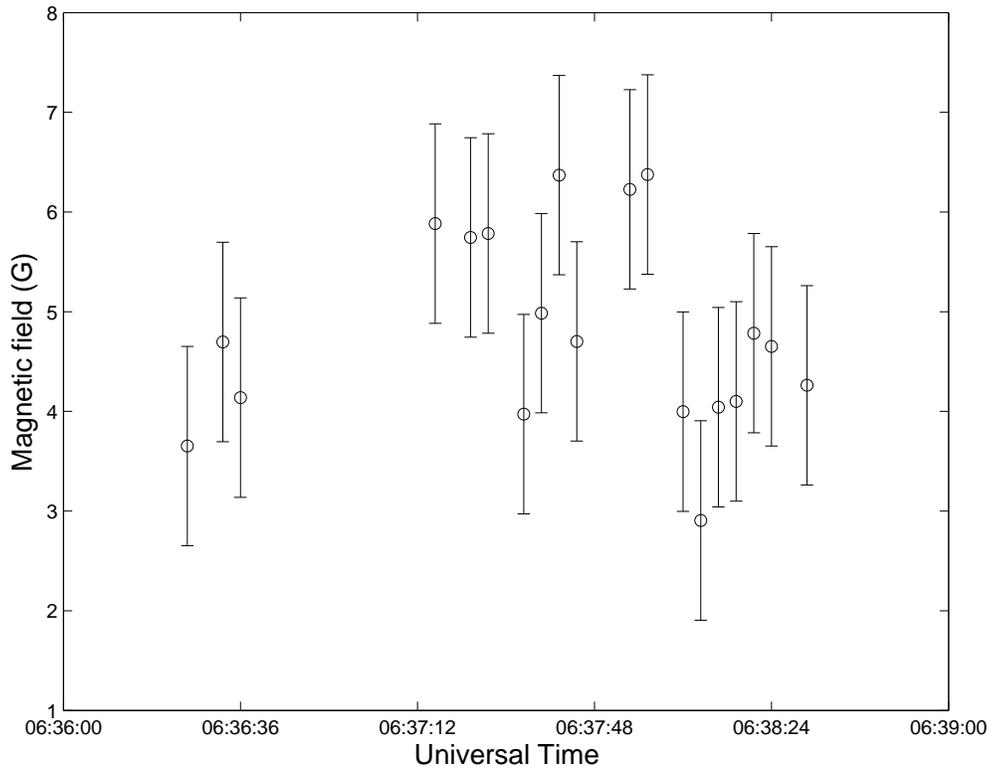}
\caption{$B$ values corresponding to the individual type III bursts in the quasi-periodic emission
in Figure \ref{fig:figure1}, based on the relationship between $B$ and $dcp$.}
\label{fig:figure5}
\end{figure}

\citet{Wild1959} pointed out
that both the fundamental (F) and the harmonic (H) type III solar
radio burst emission 
should generally be observable only for events near the center
of the solar disk. Elsewhere it should be purely harmonic emission.
This is because the F emission is more directive compared to
the H emission. 
According to \citet{Caroubalos1974}, the ground based observations of type III 
radio bursts associated with sunspot regions located $\gtrsim 70^{\arcdeg}$ either 
to the east or west of the central meridian on the Sun, are primarily H emission. 
The F component has a limiting directivity of $\pm 65^{\arcdeg}$ (from the central
meridian on the Sun) at 80 MHz \citep{Suzuki1982}. A similar result was recently 
reported by \citet{Thejappa2012} for the very low frequency solar type III radio
bursts observed in the interplanetary medium. 
Note the heliographic longitude of the sunspot regions associated
with the type III radio bursts in the present case 
are all $\gtrsim 70^{\arcdeg}$ (see Table \ref{tab:table1}).
The estimated values of
$dcp$ for all the three events in the present case are close 
to the average $dcp$ ($\approx 0.11$)
reported for the circularly polarized H component of the type III
bursts (see Table \ref{tab:table1}). For the F component, the reported 
average $dcp$ $\approx$ 0.35 \citep{Dulk1980}.
The above arguments on the directivity and the $dcp$ indicate that the 
type III bursts observed in the present case are due to H emission.
Under such circumstances the magnetic field ($B$) near the source
region of the bursts can be calculated using the following relationship
\citep{Suzuki1978,Dulk1980,Mercier1990,Reiner2007,Melrose1978,Zlotnik1981,Ramesh2010b}:
\begin{equation}
B = {{f_{p} \times dcp }\over {2.8 ~ a(\theta,\theta_{0})}}
\end{equation}
where $f_{p}$ is the plasma frequency in MHz and $B$ is in Gauss. $a(\theta,\theta_{0})$  
is a slowly varying function which depends on 
the viewing angle $\theta$ between the magnetic field component 
and the line-of-sight,
and the angular distribution $\theta_{0}$ of the Langmuir waves.  
In the present case, $\theta \approx 70^{\arcdeg}-90^{\arcdeg}$ 
(see Table \ref{tab:table1}). We assumed $f_{p}$ = 40 MHz since the observed
emission at 80 MHz is most likely the H component. 
The Langmuir waves are 
considered to be 
confined to a small range of angles, i.e. 
a cone of opening angle $\theta_{0} \approx 10^{\arcdeg}-30^{\arcdeg}$,
to the magnetic field direction for harmonic type III emission in the `o' mode 
\citep{Melrose1978,Dulk1980,Willes1997,Benz2002}. 
Assuming the average value, i.e. $\theta_{0} \approx 20^{\arcdeg}$, we find 
that $a(\theta,\theta_{0}) \approx 0.4$ 
in the aforementioned range of $\theta$ \citep{Melrose1980,Gary1982,Suzuki1985}.
The estimated magnetic field values using equation (1) for the type III events reported
in the present work are listed in Table \ref{tab:table1}. 
Figure \ref{fig:figure5} shows the $B$ values corresponding to the individual bursts
in the quasi-periodic type III burst emission observed on 20 September 2012 (Figure \ref{fig:figure1}). 
They remain approximately constant, within the error limits. We found that the $B$ values 
corresponding to the bursts observed on 2013 January 18 and 2013 March 11 also exhibit a similar trend.

\subsection {Estimate of $B$ from the observed quasi-periodicity of the type III radio bursts}

The term quasi-periodicity or pulsations in solar radio physics refers to 
the quasi-periodic amplitude variations in the time profile of the observed radio flux
at any particular observing frequency. The associated physical mechanisms 
have been classified into three categories as on date:
(1) modulation of the emission by coronal loop oscillations, 
(2) intrinsic oscillations of the emission created by oscillatory
wave-wave interactions and wave-particle interactions, and (3) modulation of
electron acceleration/injection process responsible for the emission
\citep{Aschwanden1987,Nindos2007}. Out of this, category (2)
can be ruled out in the present case since they apply primarily to
fundamental `o' mode and harmonic extra-ordinary mode (`e' mode) whereas 
the type III bursts reported in the present work correspond to harmonic `o' mode
as mentioned in Section 3.1 \citep{Aschwanden1988}.
In view of this, we have carried out 
the calculations for categories (1) and (3)
in the reminder of this section. 
Laboratory experiments and numerical simulations
have shown that the release of magnetic energy by reconnection must be considered as
a highly time-dependent process (see for eg. \citet{Pick1990}). 
According to \citet{Kliem2000}, the temporal variations in the radio 
burst flux 
are caused by modulations of the particle acceleration in a highly dynamic 
reconnection process. 
A similar model, i.e. quasi-periodic reconnection and particle injection 
was reported by \citet{Zlotnik2003} also. The authors had noted that 
even a very small ($\approx 2\%$) quasi-periodic modulation of 
the magnetic field is sufficient for periodic electron acceleration. 
Both these models 
belong to the
category (3) mentioned above and were successful in explaining
the associated observations. 
The modulation is likely to be communicated on a magnetohydrodynamic (MHD) time
scale in the acceleration region 
\citep{Tajima1987,Aschwanden1994,Kliem2000,Asai2001}. In such a case, the 
corresponding Alfv\'{e}n speed ($v_{A}$) can be estimated as: 
\begin{equation}
v_{A} \approx l/p
\end{equation}
where $p$ is the period of the quasi-periodic emission in seconds (s) and 
$l \approx$ 10000 km is the typical dimension of the region
over which the type III radio burst producing electrons are 
injected \citep{Lantos1984,Aschwanden2002}. It is possible that the
individual bursts that are temporally separated in a type III group 
are also spatially fragmented
within the same acceleration region \citep{Pick1990,Vlahos1995,Isliker1998}.
Considering category (1) where the observed quasi-periodicities are 
caused by the MHD oscillations of the associated coronal 
loop(s), the corresponding relationship for the Alfv\'{e}n speed is \citep{Roberts1984},
\begin{equation}
v_{A} = 2.6 a/p
\end{equation} 
where $a \approx$ 3500 km 
is the width of the coronal loop. According to 
\citet{Aschwanden1987}, the quasi-periodic emission observed in the upper part of 
the corona from where the meter-decameter wave radio emission originates,
paticularly those due to plasma processes as in the present case, 
are best described by the above model. \citet{Asai2001}
showed that the quasi-periodic pulsation observed by them in the microwave range
are due to the modulation of the acceleration/injection rate of the 
non thermal electrons by the oscillations in the associated coronal loop(s).
Once $v_{A}$ is known in either category (1) or (3), 
the associated magnetic field ($B$) can be calculated from the
definition of the former, i.e.:
\begin{equation}
v_{A} = 2.05 \times 10^{6}~B~N_{e}^{-1/2}
\end{equation}
$N_{e}$ is the electron density in units of $\rm cm^{-3}$
and can be estimated using the 
relationship $f_{p} = 9 \times 10^{-3} N_{e}^{1/2}$.
Note $v_{A}$ in equation (3) is in units of km/s. 
The $v_{A}$ and $B$ values estimated using equations (2) and (4) for the 
type III events reported in 
the present work are listed in Table \ref{tab:table1}. Equations (3) and
(4) result in nearly the same values for $v_{A}$ and $B$.

\section{Summary}

We have reported estimates of the coronal magnetic field using low frequency (80 MHz)
radio observations of quasi-periodic harmonic type III burst emission associated with
sunspot regions close to the solar limb. The data were obtained with the heliograph,
the polarimeter and the spectrograph at the Gauribidanur observatory.
Though several type III bursts
are observed everyday with the aforementioned suite of instruments, we selected only 
those events
with $dcp \lesssim$ 0.15 and whose associated source region is close to the limb of
the Sun. The above are the criteria to identify a harmonic type III burst
as mentioned earlier.
Also we limited ourselves to data obtained close to the transit of the Sun
over the local meridian at Gauribidanur in order to minimize errors in the source 
position due to propagation effects. We adopted two
independant approaches to estimate the magnetic field: (1) based on the relationship between 
the $dcp$ and the $B$ for harmonic type III radio burst emission, and (2) using the 
quasi-periodicity in the observed type III radio burst emission. 
Both the methods give the same result, i.e. the average $B \approx 4.2$ G.
We would like to note here that the often referred 
empirical relationship  
for the coronal magnetic field \citep{Dulk1978} predicts
$B \approx$ 3 G at 80 MHz.
We had assumed the 80 MHz plasma level to be located at radial distance 
$r \approx 1.3~\rm R_{\odot}$ in the solar atmosphere for the above
calculation \citep{Ramesh2011b}. Note \citet{Lin2004} had measured 
$B \approx$ 4 G at $r \approx 1.1~\rm R_{\odot}$ using Zeeman splitting observations
of the Fe XIII $\lambda$1075 nm coronal emission line from an active region close to the solar limb.
Assuming that $B \propto r^{-1.5}$ \citep{Dulk1978}, we find that the
above measurement indicates $B \approx$ 3.1 G $r \approx 1.3~R_{\odot}$.
The present estimates are also in reasonable agreement with that
of \citet{Ramesh2010a} who had reported $B \approx 5 \pm 1$ G at 77 MHz for radio sources associated
with the coronal streamers at $r \rm \approx 1.5~R_{\odot}$. Note the type III radio bursts 
are considered to be closely associated with the coronal streamers \citep{Kundu1983}.
We would like to add here that the estimated $v_{A}$ in the present case 
(see Table \ref{tab:table1}) are consistent with the corresponding values reported
by \citet{Gopalswamy2001,Vrsnak2002} for the active region solar corona.

\acknowledgements

It is a pleasure to thank the staff of the Gauribidanur observatory
for their help in observations, maintenance of the antenna 
and receiver systems there. We thank the referee for his/her kind 
suggestions and comments using which we were able to bring out 
the results more clearly.

\clearpage

\clearpage

\clearpage

\clearpage

\clearpage

\clearpage

\begin{deluxetable}{cccccccccc}
\tabletypesize{\scriptsize}
\tablecaption{Paramaters related to the group of type III radio bursts observed 
with the Gauribidanur facilities \label{tab:table1}}
\tablewidth{0pt}
\tablehead{
\colhead{S.No}. & \colhead{Date} & \colhead{Time} & \colhead{Period} & \colhead{Alfv\'{e}n} & 
\colhead{Sunspot} & \colhead{Viewing} & \colhead{$dcp$} 
& \multicolumn{2}{c}{Magnetic field $B$ (G)} \\
\cline{9-10}
& & \colhead{(UT)} & {$p$} & \colhead{speed $v_{A}$} & \colhead{heliographic} & \colhead{angle $\theta$} & 
& \colhead{Harmonic} & \colhead{Quasi-periodic} \\
& & & (s) & (km/s) & co-ordinates & (deg) & & \colhead{emission\tablenotemark{a}} & \colhead{emission\tablenotemark{b}} 
}
\startdata
1 & 20 Sep 2012 & 06:35-06:38 & 4.5 & 2222 & S25W69 & 71 & 0.13 & 4.7 $\pm$ 1 & 4.8 $\pm$ 0.3  \\
2 & 18 Jan 2013 & 06:51-06:57 & 6.4 & 1562 & N18W88 & 88 & 0.08 & 2.9 $\pm$ 1 & 3.4 $\pm$ 0.3  \\
3 & 11 Mar 2013 & 07:00-07:03 & 4.8 & 2083 & N10E88 & 88 & 0.14 & 5.0 $\pm$ 1 & 4.5 $\pm$ 0.3  \\
\hline
\enddata
\tablenotetext{a}{See Section 3.1 for details.}
\tablenotetext{b}{See Section 3.2 for details.}
\end{deluxetable}

\end{document}